\documentclass{aa}
\usepackage{graphics}

\begin{document}

\thesaurus{03(02.01.2;02.12.3;11.01.2;11.17.2)}

\title{Emission lines from illuminated warped accretion disks in AGN}

\author{Rumen Bachev}

\institute{Institute of Astronomy, Bulgarian Academy of Sciences, 72
Tsarigradsko Chausse Blvd., 1784 Sofia, Bulgaria 
\newline e-mail: bachevr@astro.bas.bg}

\date{Received \dots / Accepted \dots}

\titlerunning{Lines from warped accretion disks}

\authorrunning{R. Bachev}

\maketitle

\begin{abstract}
The broad line profiles (\ion{H}{$\beta$}) from a nonplanar accretion disk resulting
from reprocessing of hard X-ray radiation are studied in this paper. 
The simplest model of a disk with small inclination and a central point-like source 
of ionizing radiation is adopted. The profiles obtained depend significantly on the 
line of sight -- they are asymmetric as well as frequency-shifted in most cases. 
Since such frequency shifted asymmetric profiles are generally not observed in AGN, 
the most plausible interpretation may be that either the major part of the broad 
emission does not originate from the disk, but from different matter near the hole, 
or the disk is in general planar. The latter situation might apply to non-rotating 
black holes, for which the accretion disk is not warped. The emission lines are then 
similar to those from planar disks, i. e. almost symmetric, double-peaked and non 
frequency-shifted. Symmetric and double-peaked profiles might also be observed in 
case of an accretion disk transparent to visual light and/or X-rays.
\end{abstract}

\section{Introduction}

It is widely believed that Active Galactic Nuclei (AGN) are powered by accretion onto 
a supermassive black hole (Rees 1984, Blandford 1990, Ho 1998). This accretion probably 
operates through a geometrically thin disk, which is necessary to ensure enough radiation 
efficiency of the process (Rees 1984). Viscous release of energy in the disk, due to 
differential Keplerian rotation, is the most commonly invoked mechanism to explain the 
observed AGN luminosities, which are usually about $10^{43-45}$ $\mathrm{erg\,s}^{-1}$. 
Through this mechanism, which generates mainly visual, UV, and soft X-ray continuum, up 
to $0.30$ of the total mass-energy of the flow can be extracted (Thorne 1974). 
The observed hard X-ray radiation is probably produced by a compact source, located 
near the center (at $R=10-20 \,R_{\mathrm{G}}$, where $R_{\mathrm{G}}=G\,M_{\mathrm{BH}}\, 
c^{-2}$, $M_{\mathrm{BH}}$ is the black hole mass). The compactness of this X-ray source 
follows from its rapid variability in many objects.

A thin accretion disk in AGN could not only be the energy source but also act as a 
fluorescent screen, producing a significant part of the broad emission lines in Seyfert 1 
type nuclei due to the reprocessing of the central hard radiation (Collin-Souffrin 1987). 
Evidence for a disk reprocessing is given by reverberation mapping studies, which show a 
very small BLR radius (several light days), in a contradiction with the standard cloud 
models predictions (Osterbrock 1993 and references therein). Detailed analysis by 
Collin-Souffrin \& Dumont (1989) shows that the low ionization lines (Balmer lines, 
for instance) from the disk could be successfully modeled, as this emission depends on 
the amount of hard X-ray radiation intercepted, instead of on the ionization parameter 
which is generally unknown. 
Two types of illuminating source geometry have been proposed (Dumont \& Collin-Souffrin 
1990a): a point source, located above/below the disk plane and a diffuse hot medium, 
reflecting hard radiation to the disk surface. The line profiles from the disk are broad, 
with $\mathrm{FWHM}\simeq 10^{3-4}\,\mathrm{km\,s}^{-1}$, double-peaked in most cases, 
symmetric and non frequency shifted (Dumont \& Collin-Souffrin 1990b). 
They depend mostly on disk structure, geometry and power of the ionizing source, 
as well as on the point of view.

All these profile calculations have been performed under a default assumption that the 
disk is a planar, axisymmetric structure, illuminated uniformly. This, however, might 
not always be true. If the accreting matter falls onto a rapidly rotating (\textit{Kerr}) 
black hole and the angular momenta of the hole, and of the accreting gas, are not aligned, 
the disk formed is a nonplanar structure (a \textit{twisted} or \textit{warped} disk). 
This is the well known \textit{Bardeen-Petterson effect} (Bardeen \& Petterson 1975, 
Macdonald et al. 1986). This effect is due to the differential \textit{Lense-Thirring 
precession} ($\omega _{\mathrm{LT}} = \frac {2aG^{2}M_{\mathrm{BH}}^{2}}{c^{3}R^{3}}$, 
$a$ is the dimensionless spin parameter of the hole, 
$0\leq a\leq 1$) of orbits around the spin axis of a rotating black hole (Misner et 
al. 1973). Although $\omega _{\mathrm{LT}}$ depends on the position within the disk, 
the disk is a steady structure as the viscosity prevents its disassembling. 
Near the hole, the flow is aligned with the equatorial plane of the 
black hole, while at larger distances it is smoothly tilted to its initial orbital 
plane. The alignment radius (the \textit{Bardeen-Petterson radius}) was originally 
been assumed to be of the same order as the radial distance, where the precession 
time scale ($\sim \omega _{\mathrm{LT}}^{-1}$) is comparable to the infall time scale 
($\sim R\,V_{\mathrm{R}}^{-1}$, $V_{\mathrm{R}}$ is the radial velocity of the gas). 
When the internal hydrodynamics of the disk is fully taken into account it appears 
to be much smaller (Kumar \& Pringle 1985). 

Accretion disks spin up nonrotating, or slowly rotating, black holes because the 
angular momentum per mass unit of the accreting gas at the innermost stable orbit 
exceeds that of the hole (Misner et al. 1973; Shapiro \& Teukolsky 1983). The 
\textit{Blandford-Znajek process} (electromagnetic extraction of the black hole 
spin energy) is probably not efficient enough to reduce completely this accumulated 
spin momentum (Modersky et al. 1997, Livio et al. 1998). In other words, an AGN black 
hole is most probably fast rotating ($a\simeq 1$). As a result, the accretion disk 
in AGN should be nonplanar in most cases, as there is no reason to suppose that the 
spin momentum of the accreting mass and the black hole spin momentum are always aligned. 
Evidence for the existence of nonplanar disks in AGN has been recently found by 
Nishiura et al. (1998).Obviously, such a nonplanar geometry of the disk makes it 
possible for the central radiation to reach the outer parts, as these can be 
directly ''seen'' from the center (Petterson 1977). The covering factor of a nonplanar 
disk in case of central irradiation is close to its initial inclination angle, measured 
in $4\mathrm{\pi}$ units.

In this paper, the broad Balmer emission due to reprocessing of the central high 
energy radiation by a warped accretion disk is investigated. Here we present 
\ion{H}{$\beta$} profiles. The profiles of other strong low ionization lines 
(\ion{H}{$\alpha$}, for instance) could be slightly different because of different 
reprocessing properties of the medium. Our main purpose here is to demonstrate the 
effect of disk twisting on the line profiles, without making fits to observational 
data. 
That is why we choose the simplest model of a disk with small inclination and a 
point X-ray source located near the center. In the next sections we describe our 
method (Sect. 2) and results (Sect. 3). Discussion, comparison with the observations, 
and conclusions are given in Sect. 4

\section{Modeling line profiles}

In our model a viscous, geometrically thin, and warped -- because of the 
Bardeen-Petterson effect -- accretion disk is irradiated by a point-like 
central source. We use a cylindrical coordinate system, centered at the black 
hole (Fig. 1). The irradiating source is located at $Z_{\mathrm{S}}=10\, 
R_{\mathrm{G}}$ above the disk plane, along the black hole axis instead of 
at the exact center, as this is a more realistic situation. 
The profiles of \ion{H}{$\beta$} emission as a result of the reprocessing of 
the central hard X-ray radiation are obtained. The disk thickness is neglected 
in the computations -- it does not affect the profile shapes significantly as the 
semithickness to radius ratio is usually much less than the tilt angle. All 
relativistic corrections are also neglected, because they are small while our goal 
is to obtain qualitative results only. 
The disk is assumed to be optically thick to visual light and X-rays -- no emission 
from the lower surface (probably also illuminated) can reach an observer located 
above the X-Y plane.

\subsection{Geometry of nonplanar disks}

Since the precession velocity is small compared to the Keplerian orbital velocity 
$V_{\mathrm{K}}=c\,R_{\mathrm{G}}^{0.5}R^{-0.5}$, the disk is a stationary structure 
and can be treated as being composed of concentric rings, laying in different planes. 
Warp waves can propagate through the disk surface if $\alpha<<1$ (Papaloizou \& 
Pringle 1983), where $\alpha$ is the dimensionless viscosity parameter (Shakura \& 
Sunyaev 1973), but this is not the case for AGN accretion disks, where $\alpha$ is 
usually assumed to be 0.1-1. Each disk ring is defined by two Eulerian angles $\beta$ 
and $\gamma$ (Fig. 1) and its radius $R$.

\begin{figure}
\resizebox{\hsize}{!}{\includegraphics{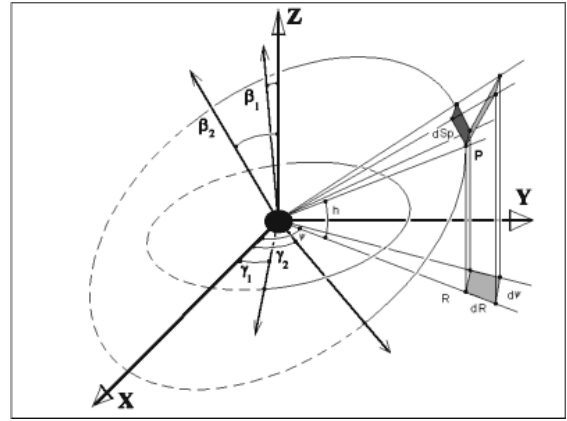}}
\caption[]{The twisted accretion disk -- a steady system of inclined rings, each 
defined by two Eulerian angles and the radial distance ($R$). The axis of the 
black hole is aligned with the Z direction. To explore the irradiation, the disk 
is divided into $\sim \, 10^{7}$ elements. Cylindrical coordinates are used to 
express the position of each element.}
\end{figure}

For a stationary twisted accretion disk, $\beta $ and $\gamma $ are slowly varying 
functions of the radial distance $R$; $\beta =\beta (R)$, $\gamma =\gamma (R)$ and 
$\beta \mathrm{|}_{\mathrm{R\rightarrow \infty}}\longrightarrow \beta _{0}$, 
$\beta \mathrm{|}_{\mathrm{R\rightarrow 0}}\longrightarrow 0$, 
$\gamma \mathrm{|}_{\mathrm{R\rightarrow \infty}}\longrightarrow 0$, 
as is originally shown by Bardeen \& Petterson (1975).
In this paper we use the steady solution of Scheuer \& Feiler (1996), derived 
following Pringle (1992). This solution, shown in Fig. 2, can be presented 
analytically by:

\begin{eqnarray}
\gamma (R) &\approx &\Gamma \,R^{-m}  \nonumber \\
\beta (R) &\approx &\beta_{0}\,\mathrm{e}^{-\gamma (R)}.
\end{eqnarray}

Here the dimensionless parameter $\Gamma \simeq 400 \,\sqrt{a\,\alpha}$ if 
the semithickness-to-radius ratio of the accretion disk is about 0.01 
(Collin-Souffrin \& Dumont 1990), $m=0.75$, and $R$ is measured in 
$R_{\mathrm{G}}$. A similar solution has been obtained by Bardeen \& Petterson 
(1975) and Hatchett at al. (1981), althoughthey did not take into account the 
internal hydrodynamics of the flow. In that case though, $\Gamma$ and $m$ are 
different and depend mostly on the radial 
velocity and the Kerr parameter ($a$), and the resulting alignment radius is 
much larger. Eqs. 1 are valid for $\alpha \simeq 0.3$ (Kumar \& Pringle 1985, 
Natarajan \& Pringle 1998), large radial distances and small initial inclination 
angles of the flow -- $\beta _{0}<15{{}^{\circ }}$.
If the initial tilt is significant, the influence of a significant amount of 
intercepted ionizing continuum on the disk structure should be taken into account 
and the disk shape should be studied numerically.

\begin{figure}
\resizebox{\hsize}{!}{\includegraphics{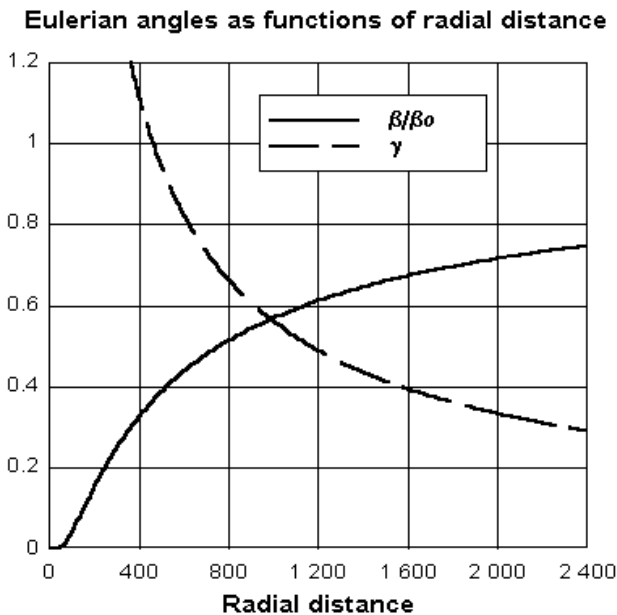}}
\caption[]{Eulerian angles $\beta$ (solid line) and $\gamma$ (dashed line) as 
functions of radial distance, according Scheuer \& Feiler (1996). Here $\Gamma=130$. 
$\beta$ is normalized to the initial tilt angle -- $\beta_{\mathrm{0}}$. $\gamma$ 
is measured in radians.}
\end{figure}

\subsection{Illumination and line profiles}

In order to explore the nonplanar disk illumination and the resulting line profiles 
we created a simple code. The input parameters are: the initial inclination angle 
$\beta _{0}$, the angles defining the line of sight, the ionizing luminosity 
$L_\mathrm{X}$, the source height $Z_{\mathrm{S}}$ and the black hole mass 
$M_{\mathrm{BH}}$.
For small inclinations of the disk, the deviation angle $h=\beta(R)\,\mathrm{sin}
(\psi-\gamma(R))$ of a disk point (P) is also small ($|h|\leq \beta \leq \beta_{0}
<0.15$) and the following approximation is valid: 
$\mathrm{sin}(h) \simeq h$; $\mathrm{cos}(h) \simeq 1$; see Fig. 1. 
The disk is divided into about $10^{7}$ elements, each defined by its dimensions 
$\mathrm{d}R$, $\mathrm{d}\psi$ and coordinates $R$, $\psi$.
An element of the grid constructed in this way absorbs an amount of the central 
hard radiation proportional to its cross section area $\mathrm{d}S_{\mathrm{P}}
(R,\psi)$ (Fig. 1), which is roughly: 
$$\mathrm{d}S_{\mathrm{P}}=R^{2}\, %
\mathrm{d}\psi \mathrm{d}R \frac{((\frac{\partial h}{\partial R})_{\mathrm{\psi}}+
\frac{Z_{\mathrm{S}}}{R^{2}})}{(1+(\frac{Z_{\mathrm{S}}}{R}-h)^{2})^{0.5}}$$
For all elements the following value is calculated:

\begin{equation}
$$\mathrm{d}L_{\mathrm{H\beta}}(R,\psi) = \frac{L_{\mathrm{X}}}{4\,\mathrm{\pi}\, %
(R^{2}+Z_{\mathrm{S}}^{2})}\, %
\mathrm{d}S_{\mathrm{P}}\, f_{\mathrm{H\beta}}(R,\psi)$$
\end{equation}

In the calculation of the total \ion{H}{$\beta$} luminosity, shadowing of the 
elements by the inner parts of the disk must be taken into account. This has 
been done as follows: The contribution of each element to the total \ion{H}{$\beta$} 
luminosity from the disk is equal to $\mathrm{d}L_{\mathrm{H\beta}}$ (Eq. [2]) 
only in case that there does not exist another element with $(h_{\mathrm{1}}
(R_{\mathrm{1}}<R, \psi)-Z_{\mathrm{S}}/R_{\mathrm{1}})>(h(R,\psi)-Z_{\mathrm{S}}/R)$ 
and is 0 in the opposite case.
$L_{\mathrm{X}}$ -- the hard X-ray luminosity, is close to the bolometric 
luminosity if the continuum extends up to $100\, \mathrm{keV}$ with a spectral 
index close to 1. $f_{\mathrm{H\beta}}(R,\psi)$ is the line response to the 
ionizing flux, i.e. it represents the probability that an absorbed X-photon 
produces an \ion{H}{$\beta$}-photon. It depends on the physical parameters of 
the emitting layers, taking into account the saturation of the lines (due to the 
limited number of line-emitting atoms), the absorbed fraction of hard radiation 
(if the column density of the absorbing gas $N<10^{25} \mathrm{cm}^{-2}$, 
not all hard radiation is absorbed), etc. It also may depend slightly on the 
incident flux -- the reprocessing is not necessarily a linear process (Eq. [2]). 
In our computations we used this dependence as given by Collin-Souffrin \& 
Dumont (1989, 1990b). In their work $f_{\mathrm{H\beta}}(R,\psi)$ is presented in a 
multifunctional form, taking into account all effects mentioned. 

To obtain the profile of the line emission from the whole disk we have to 
calculate the Doppler shift of each element, i. e. the projection ($V_{\mathrm{D}}$) 
of the Keplerian velocity onto the line of sight. 
$V_{\mathrm{D}}$ is given approximately by:
$$V_{\mathrm{D}}=V_{\mathrm{K}}(\mathrm{sin}(\psi-\psi_{\mathrm{0}})\mathrm{cos}
(h_{\mathrm{0}}) + \beta \mathrm{cos}(\psi-\gamma)\mathrm{sin}(h_{\mathrm{0}})),$$                                                                                      
where $h_{\mathrm{0}}$ and $\psi_{\mathrm{0}}$ are the angles defining the direction 
to the observer. The whole profile can be constructed by adding contributions of all 
elements, taking into account their Doppler shifts.

\section{Results}

\begin{figure}
\resizebox{\hsize}{!}{\includegraphics{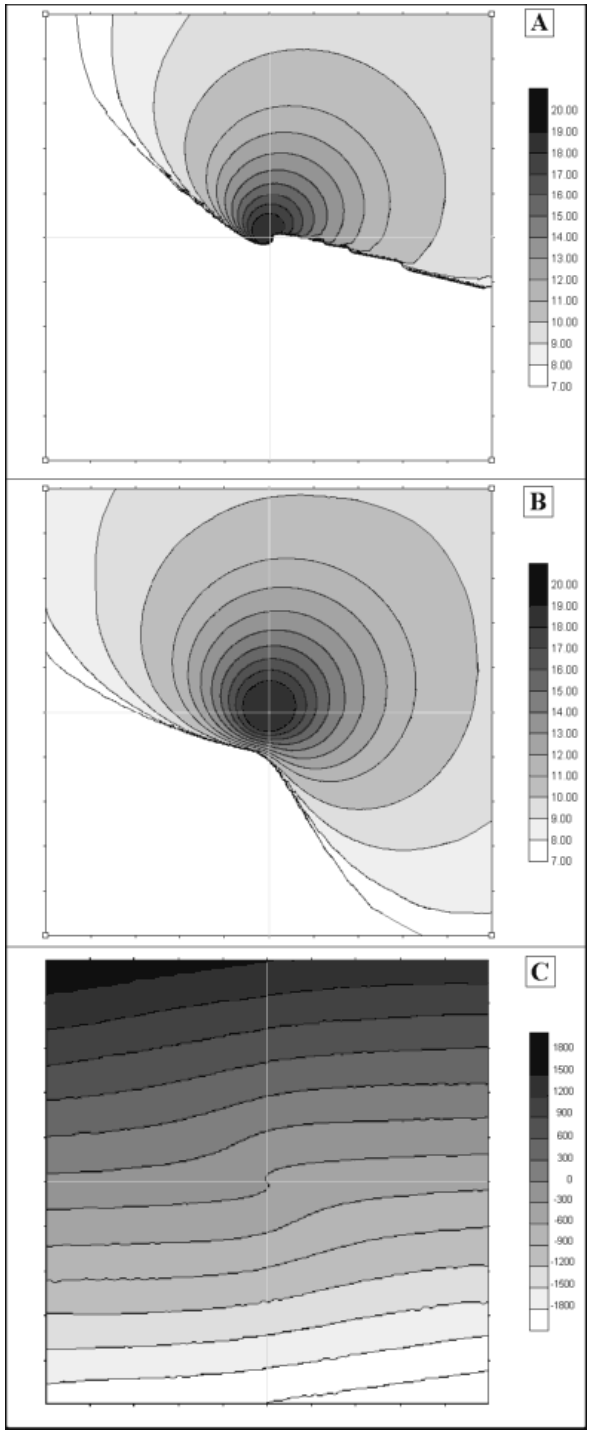}}
\caption[]{A schematic map of the \ion{H}{$\beta$} -- flux from an 
irradiated warped disk (Fig. 3a, b). The flux values are in relative 
logarithmic units. $\beta _{0} = 10{{}^{\circ }}$. 
$M_{\mathrm{BH}}=10^{8}\, M_{\mathrm{\odot}}$, $L_{\mathrm{X}}=10^{44}\, 
\mathrm{erg\,s}^{-1}$, $\Gamma=130$. Maps  for two values of 
$Z_\mathrm{S}$, $Z_\mathrm{S}=10\,R_{\mathrm{G}}$ (Fig. 3a) and 
$100\,R_{\mathrm{G}}$ (Fig. 3b) are shown. The inner ($R<10^{4}\, R_{\mathrm{G}}$) 
part of the disk is presented in the figure. 
The height (in $R_{\mathrm{G}}$) of the disk surface in respect to the X-Y plane 
is shown in Fig. 3c.}
\end{figure}

The main difference between the line emission from planar disks and warped ones 
is that in the latter case only a part of each disk ring is irradiated and 
respectively emits lines. A schemathic 2D map of the disk emission flux is given 
in Fig. 3. Profiles for different lines of sight are presented in Fig. 4. 
Here the bolometric luminosity is $10^{44}\, \mathrm{erg\,s}^{-1}$, the black hole 
mass is $10^{8}\, \mathrm{M_{\mathrm{\odot}}}$, and the initial inclination angle is 
$10{{}^{\circ }}$. 
The Kerr parameter $a=1$ and the viscosity parameter $\alpha=0.1$, so $\Gamma 
\approx 130$.

The most important result of this work is that the profiles of the disk emission 
(Fig. 4) are nonsymmetrical and red or blue frequency shifted, as expected if 
only about a half of the disk emits in the direction towards the observer (Fig. 3). 
Only for special viewing angles, these profiles could be almost symmetric, 
double or single-peaked and non frequency-shifted. In Fig. 5 the profile 
dependence of various quantities is presented. $\Gamma$ is an important parameter, 
affecting significantly the disk shape and respectively the line profiles. 
In Fig. 5a the profiles for $\Gamma=40$ (for instance $a=0.1$, $\alpha=0.1$); 
130 ($a=1$, $\alpha=0.1$) and 400 (the upper limit -- $a=1$, $\alpha=1$) are shown.
In Fig. 5b the profile dependence on the source height $Z_\mathrm{S}$ is shown. 
Increasing $Z_\mathrm{S}$, a larger part of the central disk region is irradiated 
and the profiles approach those from the planar disks -- the second, symmetrically 
displaced peak increases its intensity. We do not see any physical reason, however, 
to put the source far above the central object. 
The dependence of the profiles on the hard X-ray luminosity is also significant, 
because of the effect of the saturation of the lines, which limits the emission 
line flux at small distances (high velocities), independent
of an increase of $L_{\mathrm{X}}$. For higher $L_{\mathrm{X}}$, the lines should 
become narrower (Fig. 5c). One should keep in mind, however, that $L_{\mathrm{X}}$ 
should be $10^{43-45}\, \mathrm{erg\,s}^{-1}$ to match the observed line intensities.
Profile shapes are almost independent of the initial disk tilt (Fig. 5d). The 
$\mathrm{FWHM}$ and the shift slightly decrease with increasing tilt.
Note that the profiles are also not dependent directly on the black hole mass 
and the accretion rate (of course, the mass and the accretion rate may affect 
the incident flux at a given distance and $f_{\mathrm{H\beta}}(R,\psi)$ respectively).

Weaker emission from the opposite side of the disk may appear in case the disk is 
not fully opaque at visual wavelengths. This might be the case for the outer 
regions, where the optical depth is probably not significant if no dust is present 
(Collin-Souffrin \& Dumont 1990). Profiles  should then be double-peaked and more 
or less symmetric. Although this case is not considered here, one can easily 
reproduce such profiles. A similar situation occurs when the disk is transparent 
to the hard radiation, which is then absorbed within the whole vertical structure 
of the disk. 

\begin{figure}
\resizebox{\hsize}{!}{\includegraphics{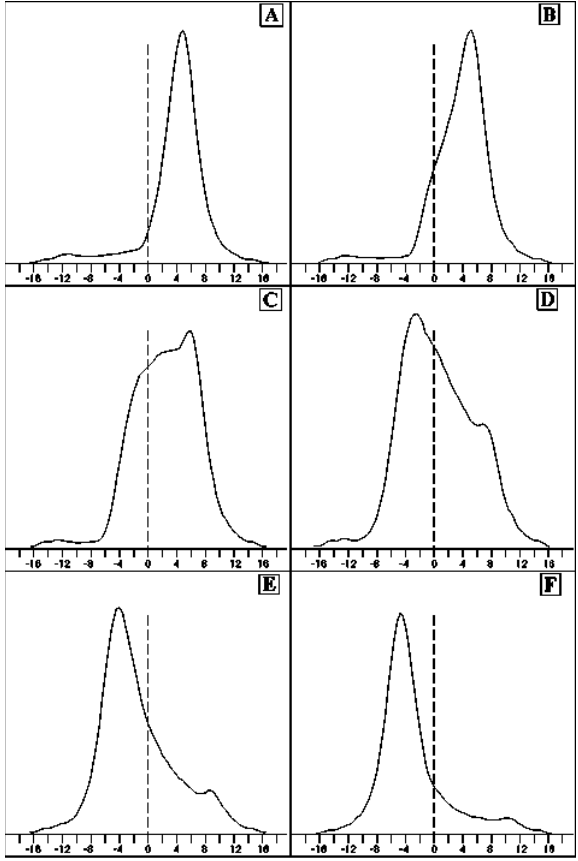}}
\caption[]{Line profiles of \ion{H}{$\beta$} emission from the disk, shown in fig. 
3a, for different viewing angles. The observer is tilted at an angle 
$h_{\mathrm{0}}=45{{}^{\circ }}$ with respect to the black hole equatorial plane. 
The azimuthal angle along the X-Y plane ($\psi_{\mathrm{0}}$) is respectively $0, 
30, .. 150{{}^{\circ }}$ -- fig. 4a..f. The inner and the outer radii of the disk 
are $10$ and $10^{4} \, R_{\mathrm{G}}$ respectively. Abscissa is in $10^{3} \, 
\mathrm{km\,s}^{-1}$.}
\end{figure}

\section{Discussion and conclusion}

The hard X-ray radiation, coming from the center, can illuminate the outer 
parts of the nonplanar disk, where optical emission lines will be emitted as 
a result of reprocessing. Profiles of such lines are modeled here (Fig. 4). 
Changing some basic parameters, such as the hard X-ray luminosity $L_{\mathrm{X}}$, 
the source height $Z_\mathrm{S}$, $\Gamma$, etc., a large variety of asymmetric 
profiles, displaced in frequency from the systemic velocity profiles can be produced.

\begin{figure}
\resizebox{\hsize}{!}{\includegraphics{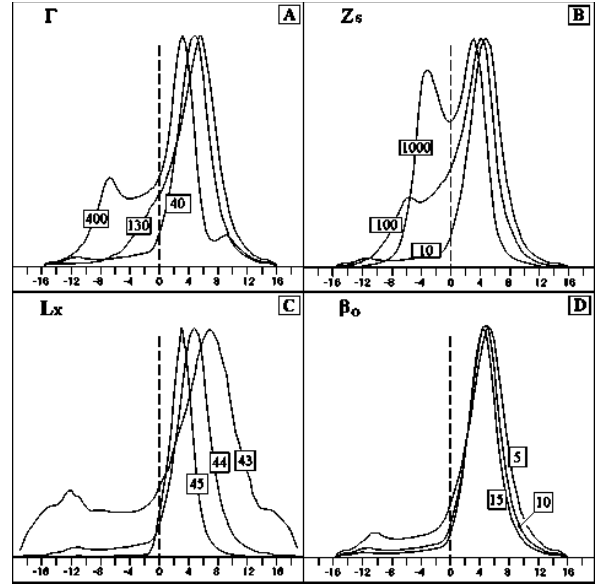}}
\caption[]{The profile dependence for different factors: the value of $\Gamma$ -- 
fig. 5a; the source height -- fig 5b ($Z_\mathrm{S}$ is measured in $R_{\mathrm{G}}$); 
the central hard luminosity -- fig. 5c ($\mathrm{log}(L_{\mathrm{X}})$, with 
$L_{\mathrm{X}}$ in $\mathrm{erg\,s}^{-1}$) and the initial tilt angle -- fig. 5d 
($\beta_{\mathrm{0}}$ is in degrees). 
$h_{\mathrm{0}}=45{{}^{\circ }}$, $\psi_{\mathrm{0}}=0{{}^{\circ }}$. 
Abscissa is in $10^{3} \, \mathrm{km\,s}^{-1}$. 
All profiles are normalized to the maximal value.}
\end{figure}

Signatures of warped disk profiles can be seen in many objects (for instance 3C 227, 
Mkn 668, 3C 390.3, etc., Eracleous \& Halpern 1994). It is possible that these 
objects contain an illuminated warped disk, but of course, other explanations can 
not be excluded. However, most of the AGN show symmetric profiles, without frequency 
shifts. On average, no more than roughly 10-15\% of the broad line emission in AGN 
can be reproduced by an irradiated nonplanar disk. We note that this value is close 
to the covering factor of such a disk. One may conclude, therefore, that the bulk of 
the line emission in AGN arises from some more or less spherical or conical structure 
(a system of clouds or star atmospheres, a jet), with a covering factor which is close 
to unity, but not from a irradiated thin disk. In that case, the presence of an inclined 
disk structure will cause only asymmetry of lines or additional peaks, displaced in 
frequency from the main profile. There are several possibilities for objects where no 
such asymmetries or displaced peaks are observed.

i) The presence of a thin disk in AGN is quite unusual. The disk may also be thick 
(slim, advection dominated -- Rees 1984; Blandford \& Begelman 1998; Chakrabarti 1998; 
Park \& Ostriker 1998).

ii) The central black hole is nonrotating and the disk is not warped. In this case, 
if the irradiation is possible, the profiles should be symmetric and double-peaked, 
which is also seldom observed, however. Similar profiles should be observed in case 
of a disk transparent to visual light and/or X-rays.

iii) The spins of the infalling gas and the hole are aligned and the disk is not 
warped. This is possible if the accreting matter is supplied from a single direction 
for a long period of time. As a result the black hole would align its spin with the 
spin of the accreting matter on relatively short time scale -- $10^{6-7} \,\mathrm{yr}$ 
(Scheuer \& Feiler 1996; Natarajan \& Pringle 1998).

iv) The disk is warped, but the illuminating source is positioned far above the disk 
plane -- at $\sim 10^{3}\, R_{\mathrm{G}}$ or more (or a huge reflecting corona with 
a significant optical depth is present). In this case a small inclination of the disk 
would not affect significantly the line profiles.

Nevertheless, many objects reveal asymmetries and displaced peaks, which could be 
successfully reproduced by a warped accretion disk. These profiles depend strongly 
on the twisting structure, the illuminating source geometry and the line of sight to 
the observer. In principle they can be modeled with high signal-to-noise spectra 
of the object by varying the model parameters. Knowledge of the twisting structure 
may allow the determination of some very important characteristics of AGN, at first 
order the spin momentum of the black hole and the viscosity parameter of the accretion 
disk. Their determination by other methods is still unreliable.

\begin{acknowledgements}
The author would like to acknowledge the kind hospitality of Kapteyn Institute, The 
Netherlands, where much of this work has been done. It is a pleasure to thank Prof. 
T. S. van Albada for his encouragement during this research, valuable comments and 
suggestions. Thanks are also due to Dr. G. T. Petrov, Dr. A. Strigachev and Dr. 
L. Slavcheva for their support and helpful comments. 
\end{acknowledgements}

\end{document}